\documentclass[onecolumn]{emulateapj}
\usepackage{longtable}
\usepackage{graphicx}
\usepackage{graphics}
\usepackage{color}
\usepackage{epsfig}
\usepackage{rotating}
\usepackage{subfigure}
\usepackage{float}
\usepackage{url}
\usepackage{hyperref}
\usepackage{breakurl}

\shortauthors{}
\shorttitle{}
\begin{document}

\title{Do we detect the galactic feedback material in X-ray observations of nearby galaxies? --- A case study of NGC~5866}

\author{Jiang-Tao Li\altaffilmark{1}} \altaffiltext{1}{Department of Astronomy, University of Michigan, 311 West Hall, 1085 S. University Ave, Ann Arbor, MI, 48109-1107, U.S.A.}


\nonumber

\begin{abstract}
One of the major sources of the X-ray emitting hot gas around galaxies is the feedback from supernovae (SNe), but most of this metal-enriched feedback material is often not directly detected in X-ray observations. This missing galactic feedback problem is extremely prominent in early-type galaxy bulges where there is little cool gas to make the Type~Ia SNe ejecta radiate at lower temperature beyond the X-ray domain. We herein present a deep \emph{Suzaku} observation of an S0 galaxy NGC~5866, which is relatively rich in molecular gas as an S0 galaxy and shows significant evidence of cool-hot gas interaction. By jointly analyzing the \emph{Suzaku} and an archival \emph{Chandra} data, we measure the Fe/O abundance ratio to be $\rm 7.63_{-5.52}^{+7.28}$ relative to solar values. This abundance ratio is much higher than those of spiral galaxies, and even among the highest ones of S0 and elliptical galaxies. NGC~5866 also simultaneously has the highest Fe/O abundance ratio and molecular gas mass among a small sample of gas-poor early-type galaxies. An estimation of the Fe budget indicates that NGC~5866 could preserve a larger than usual fraction, but far from the total amount of Fe injected by Type~Ia SNe. We also find that the hot gas temperature increases from inner to outer halos, with the inner region has a temperature of $\sim0.25\rm~keV$, clearly lower than that expected from Type~Ia SNe heating. This low temperature could be most naturally explained by additional cooling processes related to the cool-hot gas interaction as being indicated by the existence of many extraplanar dusty filaments. Our results indicate that the large cool gas content and the presence of cool-hot gas interaction in the inner region of NGC~5866 have significantly reduced the specific energy of the SN ejecta and so the velocity of the galactic outflow. The galaxy could thus preserve a considerable fraction of the metal-enriched feedback material from being blown out. 
\end{abstract}

\section{Introduction}\label{sec:Introduction}

The hot X-ray-emitting coronae around disk galaxies serve as a reservoir where galaxies acquire baryons to build up galactic disks and deposit kinetic energy and chemically-enriched materials. This X-ray emitting gas has been detected in various types of disk galaxies (e.g., \citealt{Li08,Li11,Li13a,Mineo12}), and can be globally reproduced by recent cosmological hydrodynamical simulations (\citealt{Crain10,Li14,Bogdan15}), supporting the expected origin of galactic corona from feedback close to the galaxy (\citealt{Strickland04,Li08,Li13b}), or maybe related to accretion of external gas at large galactocentric radii ($r\gtrsim50\rm~kpc$; \citealt{Anderson11,Dai12,Bogdan13}). 

The coronal properties are observed to correlate with various galaxy properties (\citealt{Li13b}). In particular, there is a tight correlation between the coronal soft X-ray luminosity and the total [core collapsed (CC)+Type~Ia] supernova (SN) energy injection rate. The X-ray radiation efficiency (the fraction of SN energy observed in the coronal X-ray emission) is quite low ($\sim0.4\%$), indicating that most of the feedback material is missing from a clear detection in soft X-ray (\citealt{Li13b}). In galaxies rich in cool gas, a large fraction of the feedback energy may be radiated through the recombination line of warm ionized gas at a lower temperature ($kT\sim10^{4-6}\rm~K$; e.g., \citealt{Agertz13}). However, in early-type galaxy bulges which are typically poor in cool gas, the missing feedback material is most likely carried out to the intergalactic space by a galactic bulge wind mainly produced by Type~Ia SNe (e.g., \citealt{Tang09,Tang10,Wang10}). 

The metal-enriched galactic feedback material is typically hot ($kT\gtrsim10^7\rm~K$) and has too low density ($n\lesssim{\rm a~few\times10^{-4}~cm^{-3}}$) to be directly detected in current X-ray observations (\citealt{Tang09,Tang10}). It is thus controversial what is the nature of the hot gas detected in soft X-ray around nearby galaxies: the real diffuse hot tenuous corona or just the mixed material at the interface of cool and hot gases (e.g., \citealt{Strickland02})? In order to understand the nature of this soft X-ray emission and to search for the missing galactic feedback material, we need to measure the metallicity of hot gas around isolated, moderate- or low-mass, early-type galaxies, which are weakly affected by the intra-cluster medium (ICM), gravitational heating, or current star formation. However, existing X-ray observations of such galaxies often reveal a low Fe (a primary product of Type~Ia SN) abundance, typically subsolar or slightly supersolar (e.g., \citealt{Humphrey06,Ji09,Konami14}), indicating that most of the SN ejecta is not detected.

We herein present a deep \emph{Suzaku} observation of NGC~5866. This galaxy locates in a low density environment, with two companions, NGC~5907 and NGC~5879, located several hundred kpc away in projection; their interactions with NGC~5866 should have negligible effects on the coronal properties. The AGN and star formation activities are both weak, unlikely contributing significantly to the X-ray emission. The edge-on orientation helps the gas-rich galactic disk to block most of the on-disk X-ray-bright point-like sources, reducing the complexity of analyzing the low-resolution \emph{Suzaku} data. Based on our preliminary study of NGC~5866 using \emph{Chandra} and multi-wavelength data \citep{Li09}, we found it has high Fe/O abundance ratio as a disk galaxy \citep{Li13b}, thus well-suited for studying Type~Ia SN metal enrichment. However, the metal abundances are poorly constrained with \emph{Chandra} data only, so we proposed deep \emph{Suzaku} observation of this galaxy to collect more photons ($\sim100\rm~ks$; Cycle~6; PI: Li J. T.). 

In this paper, we jointly analyze the \emph{Suzaku} and \emph{Chandra} data, and further compare this galaxy to other galaxies and discuss how the new observation could help us to understand the Type~Ia SN feedback processes. The paper is organized as follows: In \S\ref{sec:Data}, we describe the calibration of the \emph{Suzaku} data and the joint analysis of the \emph{Suzaku} and \emph{Chandra} spectra. In \S\ref{sec:Discussion}, we compare our X-ray measurements of NGC~5866 to similar measurements of other galaxies and further discuss the results and their implications. Out major results and conclusions are summarized in \S\ref{sec:Summary}. Errors are quoted at 90\% confidence interval throughout the paper, unless specifically noted.

\section{Data Reduction and Analysis}\label{sec:Data}

\subsection{Suzaku data reduction}\label{subsec:DataSuzaku}

\begin{figure}[!h]
\begin{center}
\epsfig{figure=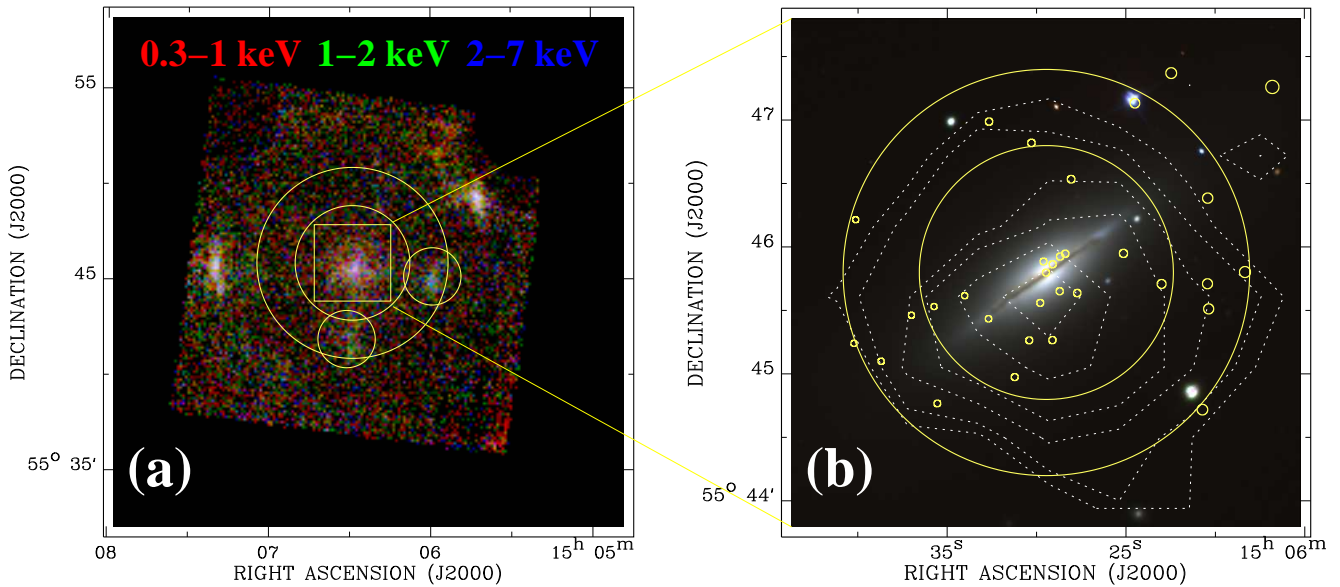,width=1.0\textwidth,angle=0, clip=}
\caption{(a) \emph{Suzaku} tri-color (red: 0.3-1~keV; green: 1-2~keV; blue: 2-7~keV) counts image of NGC~5866 and the surrounding regions. The annulus marks the region used to extract the sky background spectrum, after removing the sources in the two smaller circles. The small box marks the central $4^\prime\times4^\prime$ region centered at the nucleus of NGC~5866, and is also the FOV of panel~(b). (b) \emph{Suzaku} background-subtracted and exposure-corrected 0.3-1~keV contours overlaid on the SDSS dr7 tri-color images (red: i-band; green: r-band; blue: g-band). The inner circle has a radius of $1^\prime$ and is used to extracted the inner halo spectrum, while the outer circle has a radius of $1.6^\prime$ and the annulus between these two circles is used to extract the outer halo spectrum. The small circles mark the location and size (96\% PSF energy enclosed radius) of the \emph{Chandra} detected point-like sources (\citealt{Li09}).}\label{fig:image}
\end{center}
\end{figure}

\emph{Suzaku} observation of NGC~5866 was taken on 2011-05-20 with a total exposure time of 101.8~ks. We reprocess the XIS data using the ftool \emph{aepipeline} and CALDB files of version 2014 October 1. We identify the high background flare time intervals using a $3~\sigma$ clip with a lightcurve binned at 200~s. We then remove these background flares and obtain an effective exposure time of 100.8, 80.9, 100.2~ks for XIS0, XIS1, XIS3, respectively. We generate the XIS non-X-ray background spectra and images using the ftool \emph{xisnxbgen}. In imaging analysis, the vignetting is corrected by simulating an event file using \emph{xissim}. The images are rescaled with bin=8 (pixel size=$8.3^{\prime\prime}$), trimmed to remove low exposure regions, flat fielded, and exposure corrected. The resultant tri-color images in 0.3-1~keV, 1-2~keV, and 2-7~keV are shown in Fig.~\ref{fig:image}a, while the 0.3-1~keV contours in the central region are presented in Fig.~\ref{fig:image}b.

\emph{Suzaku} XRT has a typical size of the point spread function (PSF) of $\sim2^\prime$. As shown in Fig.~\ref{fig:image}a, except for the emission from NGC~5866, the other bright X-ray sources (the two sources at the edge of the CCD and the two fainter sources marked with the circles) are all background sources also identified by the \emph{Chandra} (\S\ref{subsec:DataChandra}; \citealt{Li09}). 

We define the regions used for extracting sky background and source spectra as shown in Fig.~\ref{fig:image}a,b. The smaller circle in Fig.~\ref{fig:image}b has a diameter of $2^\prime$, matching the XRT PSF size; we herein call it the ``inner halo'' region. The larger circle in Fig.~\ref{fig:image}b has a diameter of $3.2^\prime$; we define the annulus between these two circles as the ``outer halo'' region, outside which there is no significant diffuse X-ray emission. The large annulus in Fig.~\ref{fig:image}a (with diameters of $6^\prime-10^\prime$), after removing the two bright background sources, are used to extract the sky background spectrum.

We extract the XIS spectra and the corresponding response files of the source and sky background regions using the ftool \emph{xselect}. The non-X-ray instrumental background spectra are also extracted from the same regions. There is a weak line feature at $\sim0.62\rm~keV$ of the XIS1 spectra, which is not detected in XIS0/XIS3 or the \emph{Chandra} spectra. This feature is not common in thermal plasma spectrum. We do not find any significant time variation of it. The feature also presents over the entire XIS1 field of view. It is thus a background feature most likely caused by \emph{quiescent} solar wind charge exchange (e.g., \citealt{Yoshitake13}). We will use an additional \emph{Gauss} model to account for it in the following spectral analysis.

\subsection{Archival Chandra data}\label{subsec:DataChandra}

\emph{Suzaku} has too low angular resolution to individually detect X-ray bright point-like sources, such as AGNs or X-ray binaries, within the optical extent of NGC~5866. Unlike X-ray faint stellar sources, the contribution from these X-ray bright sources is quite uncertain, and could be the major uncertainty in characterizing the truly hot gas emission. We therefore need a joint analysis of the \emph{Chandra} and \emph{Suzaku} data to better decompose the X-ray emission from hot gas and stellar sources and/or AGN.

The archival \emph{Chandra} ACIS-S observation of NGC~5866 is presented in \citet{Li09,Li11,Li13a}. It has an effective exposure time of $\sim24.3\rm~ks$. We herein directly use the calibrated data, the diffuse X-ray images, and the detected point-like sources (Fig.~\ref{fig:image}b) from \citet{Li13a}. The \emph{Chandra} spectra are extracted from the same regions as the \emph{Suzaku} spectra. We do not remove the point sources from the diffuse X-ray emission, in order to be consistent with the \emph{Suzaku} spectra. Instead, we extract an accumulated \emph{Chandra} spectrum of all the point sources detected within the inner halo region (point sources in the outer halo have too few counts to construct a spectrum). This spectrum will be jointly analyzed with the \emph{Chandra}/\emph{Suzaku} diffuse X-ray spectra extracted from the inner halo.

\subsection{Spectral analysis}\label{subsec:SpecAnalysis}

In general, the X-ray spectra of the inner and outer halos are comprised of several different components. We characterize the hot gas emission with an optically thin ionization equilibrium thermal plasma model VAPEC. The oxygen and iron abundances are set free, while the abundances of other elements are set to the solar values. In order to better characterize the hot gas emission, we also need to quantify the various stellar contributions. With the archival \emph{Chandra} data, we reach a point source detection limit of $\sim4\times10^{37}\rm~ergs~s^{-1}$ in 0.5-8~keV. The accumulated spectrum of all the detected point sources within the inner halo region can be well modeled with a power law, which has a photon index of $\Gamma=1.4$ and an absorbing column density slightly larger than the Galactic foreground value of $N_{\rm H}\sim1.46\times10^{20}\rm~cm^{-2}$.  Below the detection limit, the unresolved stellar source emission is dominated by the emission from cataclysmic variables (CVs) and coronal active binaries (ABs), as well as the emission from unresolved low mass X-ray binaries (LMXBs) and the residual photons spilling out of the discrete source removal circles. Similar as \citet{Li09}, we model this component with a 0.5~keV MEKAL model (optically thin thermal plasma) and a $\Gamma=1.9$ power law model. Both components are well calibrated and their normalizations are scaled with the stellar mass (\citealt{Revnivtsev08}). There is still a small residual at $\sim0.62\rm~keV$ after subtracting the sky background. This feature is only presented in the XIS1 spectrum and is most likely a residual feature produced by the \emph{quiescent} solar wind charge exchange. We therefore add into the XIS1 spectrum a narrow gauss model with fixed line width of $10^{-4}\rm~keV$ to account for it.

\begin{figure}[!h]
\begin{center}
\epsfig{figure=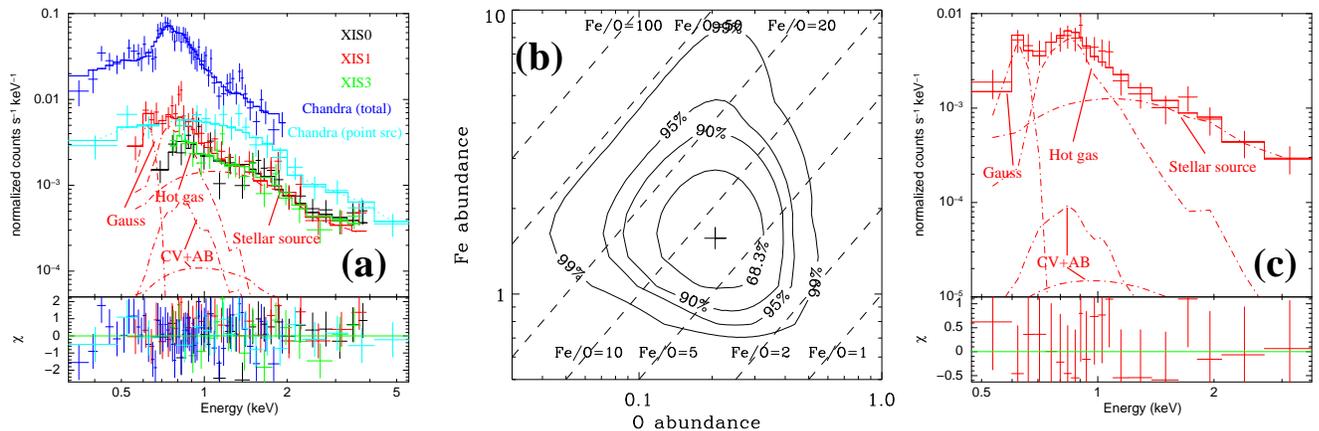,width=1.0\textwidth,angle=0, clip=}
\caption{(a) Joint analysis of the \emph{Suzaku} XIS0 (black), XIS1 (red), XIS3 (green), \emph{Chandra} total spectrum (blue), and \emph{Chandra} spectrum of point-like sources (cyan) from the ``inner halo'' region. Different model components of the XIS0 spectrum, which has the best counting statistic, are also plotted as red dash-dotted curves. (b) Confidence range of the O and Fe abundances of the model in panel~(a) [contours at 68.3\% ($1~\sigma$), 90\%, 95\%, and 99\% confidence levels]. The plus sign marks the best-fit values. The dashed lines mark different values of Fe/O abundance ratios. (c) XIS0 spectrum of the ``outer halo'' region, with different model components marked with dash-dotted curves. The abundances of O and Fe, and the power law photon index are all fixed at the values of the ``inner halo''.}\label{fig:spec}
\end{center}
\end{figure}

\begin{table}[]{}
\vspace{-0.in}
\begin{center}
\small\caption{Hot gas properties from spectral analysis}
\vspace{-0.0in}
\begin{tabular}{lcc}
\hline \hline
Parameter        & Inner halo      & Outer halo \\
\hline
$\chi^2/d.o.f.$  & $121.00/135$ & $6.59/15$ \\
$kT\rm~(keV)$ & $0.248_{-0.027}^{+0.026}$ & $0.66_{-0.11}^{+0.16}$ \\
O abundance (solar) & $0.21_{-0.10}^{+0.13}$ & - \\
Fe abundance (solar) & $1.57_{-0.58}^{+1.32}$ & - \\
$EM$ ($\rm cm^{-6}kpc^3$) & $0.053_{-0.012}^{+0.015}$ & $0.015\pm0.004$ \\
\hline \hline
\end{tabular}\label{table:SpecFitResult}
\end{center}
\vspace{-0.0in} \footnotesize Spectral analysis results of the ``inner halo'' region is obtained by jointly fitting the \emph{Suzaku} and \emph{Chandra} spectra (in total 10 free parameters). Spectral analysis results of the ``outer halo'' region is obtained by fitting the \emph{Suzaku} XIS1 spectrum with a model in which the absorption column density and photon index of the power law, as well as the O and Fe abundances of the hot gas, are all fixed at the values of the inner halo (in total 5 free parameters). See text for the detail description of the spectral models and Fig.~\ref{fig:spec}a,c for the fitted spectra. All the errors quoted in this table are statistical only, at 90\% confidence level. $EM$ is the emission measure of the VAPEC component.
\end{table}

The \emph{Suzaku} (XIS0, XIS1, XIS3) and \emph{Chandra} (total emission and detected point source emission) spectra are jointly fitted, with most of the model parameters fixed. Fitting the absorption column density ($N_{\rm H}$) of the thermal component results in an even lower $N_{\rm H}$ than the foreground value and comparable values of other parameters and $\chi^2/d.o.f.$. We therefore fix it at the foreground value of $N_{\rm H}\sim1.46\times10^{20}\rm~cm^{-2}$. The only free parameters are the temperature, O and Fe abundances, normalization (or flux) of the hot gas, the absorption column density, power law photon index and normalization of the \emph{Chandra}-detected point sources, as well as the centroid energy and normalization of the gaussian line added to the XIS1 spectrum. The joint analysis gives us much better counting statistic than presented in previous works (\citealt{Li09,Li11,Li13a}). The best-fitted spectral model parameters are summarized in Table~\ref{table:SpecFitResult} and the fitted spectra are presented in Fig.~\ref{fig:spec}a. One of our primary goal in this paper is to better determine the Fe/O abundance ratio. We plot the confidence range of Fe and O abundances in Fig.~\ref{fig:spec}b to further examine how reliable is our measurement. It is highly reliable (at $\gtrsim90\%$ confidence level) that \emph{the hot coronal gas in the inner halo of NGC~5866 has a super-solar Fe/O abundance ratio} ($\rm Fe/O=7.63_{-5.52}^{+7.28}$).

\begin{figure}[!h]
\begin{center}
\epsfig{figure=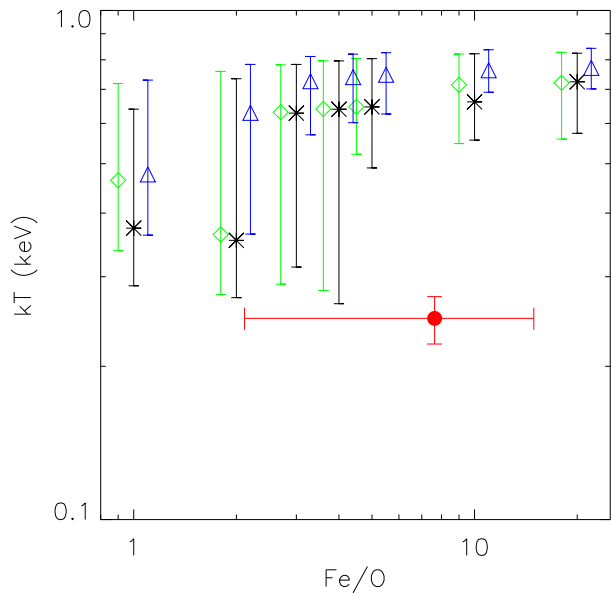,width=0.5\textwidth,angle=0, clip=}
\caption{This figure shows how the assumptions on the power law component and the different values of Fe/O ratio may affect the measurement of the hot gas temperature ($kT$) in the ``outer halo''. The black, green, and blue symbols are for models 1-3 as described in the text. They are at the same Fe/O value marked by the black symbols, but the green and blue symbols have been shifted a little along the Fe/O axis to clarify. The red dot and error bars show the measurement and 90\% confidence error of the ``inner halo'' region.}\label{fig:kTcompare}
\end{center}
\end{figure}

For the outer halo, only the XIS1 spectrum has enough counts for spectral analysis. We fit this spectrum with the same model as for the inner halo. However, the O and Fe abundances of the hot gas are poorly constrained. We then fix the abundances at the values of the inner halo, in order to estimate the hot gas temperature. The resultant $\chi^2/d.o.f.$ is too low ($\sim0.44$; Table~\ref{table:SpecFitResult}), indicating that there may be too many free parameters (five, temperature and normalization of the thermal plasma, normalization of the power law component, as well as the centroid energy and normalization of the gaussian line) and large systematic uncertainties. 

The largest systematic uncertainty which could highly affect the characterization of the hot gas emission may be the presence of the poorly constrained power law component. As the most important parameter which can be determined in the outer halo is the hot gas temperature, we further examine how the assumption of the power law model may affect the measurement of the temperature. We fit the spectrum in three different ways: (1) the absorption column density and photon index of the power law are fixed at the values of the inner halo; (2) set a lower limit on the absorption column density (at the Galactic foreground value), but free the photon index; (3) remove the power law component and fit the spectrum only in 0.3-1.5~keV, where the X-ray emission is dominated by the thermal component. In each model, we also test how the results are affected by different assumed values of the Fe/O abundance ratio (change the Fe abundance while fix the O abundance at 0.2~solar). The results are shown in Fig.~\ref{fig:kTcompare}, in comparison with the temperature of the inner halo. \emph{It is obvious that in any models, the hot gas temperature of the outer halo is higher than that of the inner halo.}

\subsection{Systematic uncertainties of the spectral analysis for the inner halo}\label{subsec:SystematicUncertainties}

The errors quoted in Table~\ref{table:SpecFitResult} only include the statistical uncertainties of spectral fitting. We herein discuss some systematic uncertainties of the temperature and abundance measurements in the inner halo. 

The contribution of the CV+AB component can exhibit a factor of $\sim1.7$ variation in different galaxies (e.g., \citealt{Boroson11}). We have adopted a relatively high $L_{\rm X,CV+AB}/M_*$ ratio from \citet{Revnivtsev08}, which may affect the measured temperature and metallicity. If we adopt a CV+AB model which has only $\sim50\%$ of the flux of the current model, the measured temperature and metallicity are only slightly different: $kT=0.255_{-0.027}^{+0.026}$, O abundance$=0.20_{-0.09}^{+0.12}$, Fe abundance$=1.44_{-0.55}^{+0.99}$. Therefore, the small uncertainty in the CV+AB component could not significantly affect our results.

We have assumed a single-temperature model of the thermal plasma, which is in general true for early-type disk galaxies like NGC~5866 (e.g., \citealt{Li11,Li13a}). However, it is still possible that the hot gas has a more complicated thermal structure. The low best-fit temperature of the hot gas indicates that it is most likely produced in the interface between the cool gas and a volume-filled hot tenuous galactic outflow (see the discussions in \S\ref{subsec:CoolGasEffect}). We further discuss how the presence of a high-temperature component representing the galactic outflow may affect our measurements of the metallicity of the low-temperature hot gas. The properties of this galactic outflow component is poorly known observationally. We therefore adopt the gas properties inferred from numerical simulations of a galactic bulge without much cool gas (\citealt{Tang09,Tang10}). According to these simulations, the hot gas temperature has a broad distribution, but the emission measure or radiative cooling coefficients typically peak at $\sim1\rm~keV$. We add a VAPEC model with a fixed temperature of $1\rm~keV$ to the model of the inner halo. The quality of the spectra do not allow us to measure the metallicity of the two components separately. We therefore simply link the O and Fe abundances to those of the low-temperature component. The best-fit model indicates an even lower temperature of the low-temperature component ($\sim0.18\rm~keV$), and an even higher Fe metallicity of $\gtrsim10~\rm solar$, further strengthening our arguments for the low-temperature and high-Fe-abundance of the hot gas in the inner region of NGC~5866.

The applied index of the power law component ($\sim1.4$), as determined with the \emph{Chandra} resolved point sources, is probably too low for the LMXBs in the galactic bulge. In nearby early-type galaxies, this index typically span in a range of 1.4-1.8, with a best fit value of 1.56 \citep{Irwin03}. We test how the value of this power law index may affect our results by fixing it at 1.8. The resultant best-fit temperature and metallicity of the inner region are: $kT=0.232_{-0.033}^{+0.030}$, O abundance$=0.20_{-0.09}^{+0.13}$, and Fe abundance$=2.04_{-0.82}^{+2.96}$, consistent with the results obtained by directly fitting this power law index.

Based on the discussions on the systematic uncertainties of the spectral analysis for the inner halo, we conclude that NGC~5866 has a significantly super-solar Fe/O abundance ratio and a low hot gas temperature of $\lesssim0.25\rm~keV$ in the inner region.

\section{Discussion}\label{sec:Discussion}

\subsection{Metallicity of the hot halo gas}\label{subsec:HotGasMetallicity}

\begin{figure}[!h]
\begin{center}
\epsfig{figure=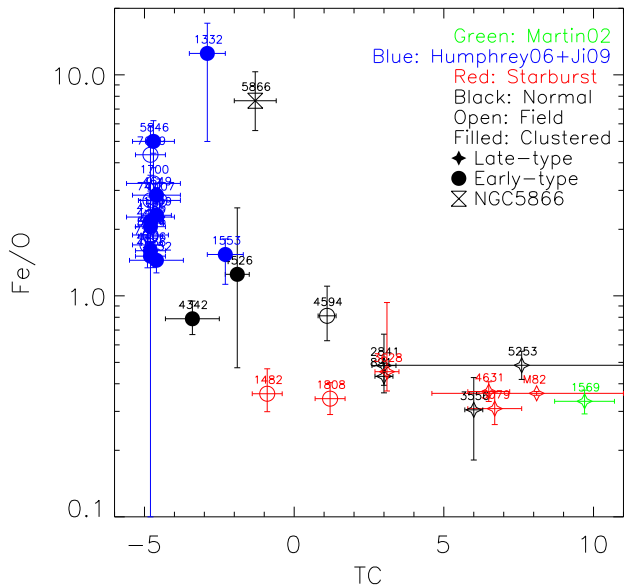,width=0.5\textwidth,angle=0, clip=}
\caption{Fe/O abundance ratio v.s. the morphological type code (TC) of NGC~5866 (black hourglass), compared to the X-ray measurements from \citet{Li13a,Li13b} (red and black symbols), \citet{Martin02} (green diamond), \citet{Humphrey06} and \citet{Ji09} (blue symbols). The error bars of the Fe/O ratio are plotted at $1~\sigma$ confidence level to clarify.}\label{fig:FeORatioTC}
\end{center}
\end{figure}

\citet{Li13b} found an anti-correlation between the hot gas Fe/O abundance ratio and the morphological type code (TC) of the galaxy, indicating that Type~Ia SN has a larger contribution in metal enrichment in earlier-type galaxies (smaller TC). We herein compare the hot gas metallicity of NGC~5866 to an updated version of the TC-Fe/O diagram presented in \citet{Li13b}. As shown in Fig.~\ref{fig:FeORatioTC}, \emph{NGC~5866 is clearly richer in Fe than other spiral and S0 ($TC\sim-3-0$) galaxies, and even many of the earlier-type elliptical galaxies.} Since most of the metal-enriched material in gas-poor S0 or elliptical galaxies is thought to be carried out by a galactic scale bulge wind or outflow (e.g., \citealt{Tang09,Tang10,Li11}), this high Fe abundance means there is a larger than usual fraction of Type~Ia SN ejecta detected in soft X-ray in the inner halo of NGC~5866.

We estimate the total mass of Fe detected in NGC~5866. Assuming spherical symmetry of the X-ray emitting regions, we could estimate the density of hot gas from the spectral fitting parameters. The densities of the inner and outer halos are $\sim1.2\times10^{-2}f^{-1/2}\rm~cm^{-3}$ and $\sim3.6\times10^{-3}f^{-1/2}\rm~cm^{-3}$, respectively, where $f$ is the poorly known volume filling factor. The total mass of hot gas contained in the inner and outer halos are both $\sim10^8f^{1/2}\rm~M_\odot$. As the metallicity in the outer halo cannot be well constrained, we only estimate the total Fe mass contained in the inner halo, which is $\sim2\times10^5f^{1/2}\rm~M_\odot$.

We further estimate the expected metal (Fe) budget in NGC~5866. In isolated gas-poor elliptical or S0 galaxies, the major source of gas is from stellar mass loss while the major source of metal is from Type~Ia SNe. Adopting a Type~Ia SN rate of $4.4\times10^{-4}\rm~SN~yr^{-1}/10^{10}~M_\odot$ (\citealt{Mannucci05}) and the stellar mass of $4.5\times10^{10}\rm~M_\odot$ in the inner halo (estimated from the K-band luminosity), we could obtain a total Type~Ia SN rate of $2\times10^{-3}\rm~SN~yr^{-1}$. If no SN ejecta can be confined around the galactic bulge, i.e., they are carried out by a galactic outflow, we would expect the detected metal is only produced within the dynamical cross timescale of the outflow. Assuming a typical outflow velocity of $\sim300\rm~km~s^{-1}$ from numerical simulations of a similar gas-poor galactic bulge (\citealt{Tang09}), the radius of $1^\prime$ of the inner halo gives a dynamical cross timescale of $t_{dyn}\sim1.5\times10^7\rm~yr$. Within this timescale, there are $\sim2\times10^4\rm~M_\odot$ Fe injected into the ISM by Type~Ia SNe (assuming each Type~Ia SN eject $\sim0.7\rm~M_\odot$ Fe; \citealt{Nomoto84}), \emph{about one order of magnitude smaller than what we have detected in the inner halo}, assuming the volume filling factor $f\sim1$. On the other hand, if most of the SN ejecta can be confined around NGC~5866, the total amount of metal injected by SNe should be calculated within a timescale since the last major starburst stage. This timescale can be estimated with the globular cluster (GC) specific frequency. We herein adopt the age after last starburst of NGC~5866 from \citet{Li09}: $t_{age}\sim8\rm~Gyr$. Within this timescale, the total Fe injected by Type~Ia SNe is $\sim10^7\rm~M_\odot$, \emph{about 50 times higher than what we have detected.}

The most uncertain parameter in the above estimation is the volume filling factor $f$. In starburst galaxies, the volume filling factor is often extremely low (e.g., $\lesssim2\%$) due to the strong interaction between the fast galactic superwind and the large amount of highly-structured cool gas (e.g., \citealt{Strickland00}). In star formation inactive galaxies such as NGC~5866, it is unlikely that $f$ is such extreme, because the amount of cool gas is much less and there is much weaker cool-hot gas interaction. We do not have a direct observational constraint on $f$. However, even if we adopt the extreme volume filling factor of starburst galaxies, the estimated total Fe mass is still consistent with or slightly larger than the Type~Ia SNe products within the dynamical cross timescale of the outflow. Therefore, we conclude that Type~Ia SNe should be sufficient to explain the observed metallicity of the hot gas in NGC~5866. The relatively larger Fe abundance of NGC~5866 compared to other elliptical or S0 galaxies indicate that NGC~5866 is able to keep a larger fraction of the SNe ejecta surrounding the galaxy, although most of the metals are still blown out, as indicated by the large difference between the observed and expected metals during the lifetime of the galaxy.

\subsection{The effect of cool gas}\label{subsec:CoolGasEffect}

\begin{figure}[!h]
\begin{center}
\epsfig{figure=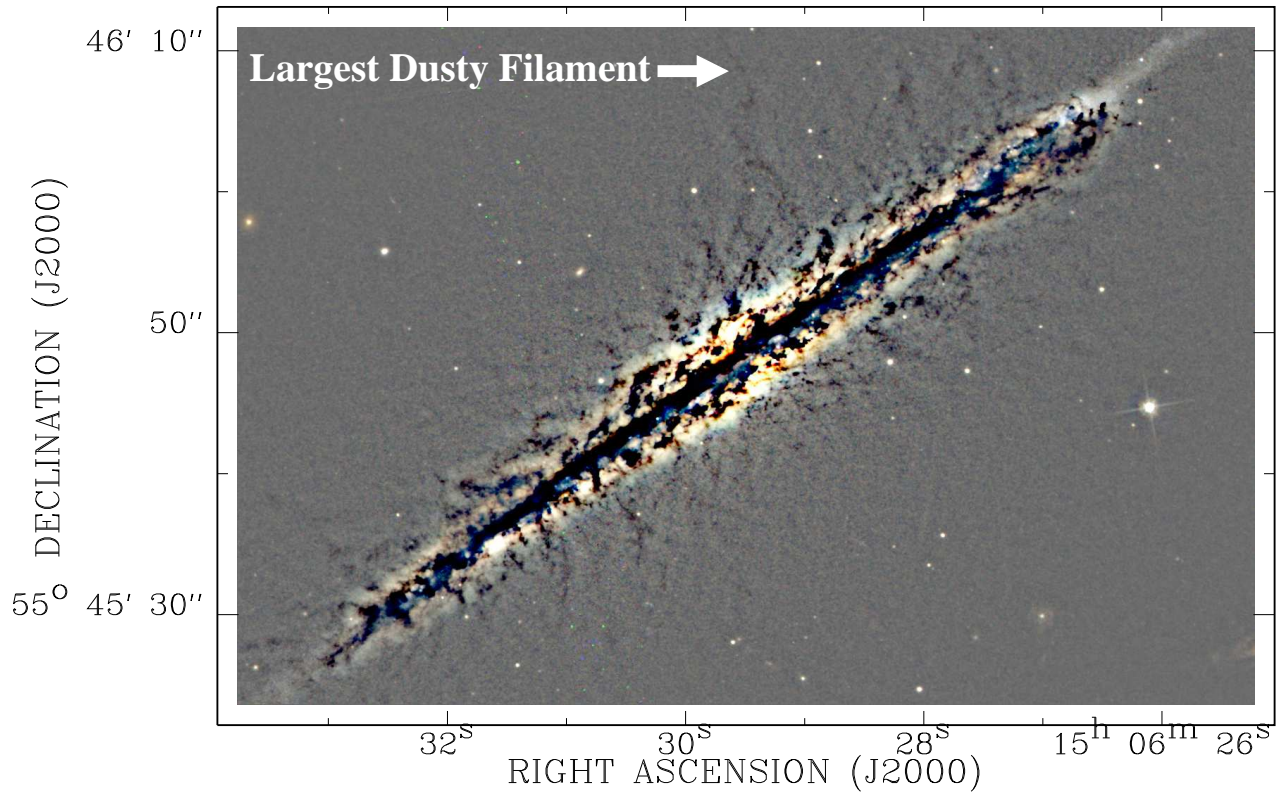,width=1.0\textwidth,angle=0, clip=}
\caption{The unsharp-masked \emph{HST} tri-color images (red: F625W; green: F555W; blue: F435W) of the central $1.2^\prime\times0.8^\prime$ of NGC~5866, showing the dusty (dark) extraplanar filaments. The arrow marks one of the largest dusty filaments, which is $\sim20^{\prime\prime}$ in length perpendicular to the galactic disk.}\label{fig:HSTfilaments}
\end{center}
\end{figure}

The perfect edge-on orientation of NGC~5866 allows us to search for dusty filaments through absorption features on the optical images, which is the best evidence for the presence of extraplanar cool gas in such gas-poor galaxies (e.g., \citealt{Howk99}). We present high resolution unsharp-masked \emph{HST} images of NGC~5866 in Fig.~\ref{fig:HSTfilaments}, which show prominent dusty filaments extended perpendicular to the galactic disk. The largest filaments extend as far as $\sim20^{\prime\prime}$ ($\sim1.5\rm~kpc$) above the disk (marked by the white arrow in Fig.~\ref{fig:HSTfilaments}). \citet{Li09} found that most of these dusty filaments, especially those close to the galactic disk, could be produced by single SN. Therefore, \emph{Type~Ia SN is energetically sufficient to power the cool-hot gas interaction in this SF inactive S0 galaxy.}

\begin{figure}[!h]
\begin{center}
\epsfig{figure=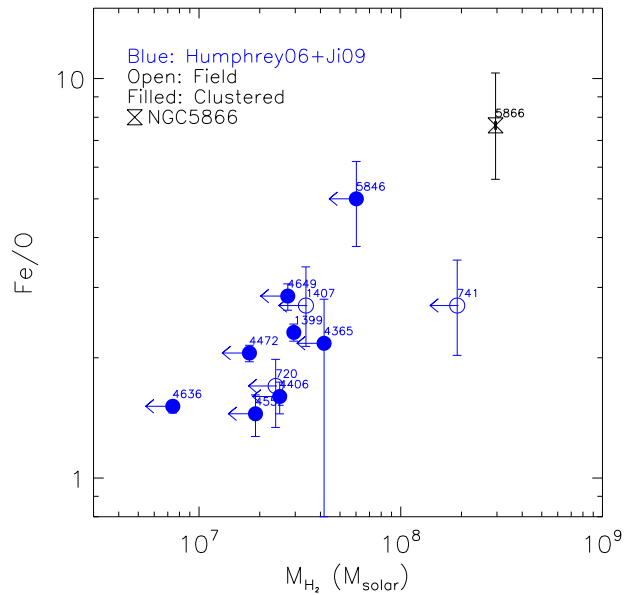,width=0.5\textwidth,angle=0, clip=}
\caption{Fe/O abundance ratio v.s. the molecular gas mass ($M_{\rm H_2}$) of NGC~5866 compared to elliptical galaxies from \citet{Humphrey06} and \citet{Ji09}. Except for NGC~5866 and NGC~1399, all the other measurements of $M_{\rm H_2}$ are upper limits ($M_{\rm H_2}$ and the references are summarized in Table~\ref{table:EarlyTypeCOAbun}). The error bars of Fe/O ratio are plotted at $1~\sigma$ confidence level to clarify.}\label{fig:FeORatioMH2}
\end{center}
\end{figure}

We further investigate the role of cool gas on the metal enrichment by comparing the Fe/O abundance ratio and the total molecular gas mass ($M_{\rm H_2}$) of a small sample of gas-poor early-type galaxies (Table~\ref{table:EarlyTypeCOAbun}). Most of these galaxies are too gas-poor so the listed $M_{\rm H_2}$ are upper limits, except for NGC~5866 and NGC~1399. It is therefore difficult to argue for a clear positive correlation in Fig.~\ref{fig:FeORatioMH2}. Nevertheless, it is clear that \emph{NGC~5866 is the most gas-rich galaxy in this sample and shows the highest gas-phase Fe/O abundance ratio}, possibly indicating that its high cold gas content may play a role in retaining the metal-enriched SN ejecta. 

\begin{table}[]{}
\vspace{-0.in}
\begin{center}
\small\caption{Hot gas metallicity and molecular gas mass of early-type galaxies}
\vspace{-0.0in}
\begin{tabular}{lccccccccccccccc}
\hline \hline
Name        & Fe/O     & Fe/O ($4^\prime$) & $\log M_{\rm H_2}/M_\odot$ & $M_{\rm H_2}$ ref. \\
\hline
NGC~5866 &  $7.63_{-5.52}^{+7.28}$ & - & $8.47\pm0.01$ & Young11 \\
NGC~720 & $1.70_{-0.98}^{+0.76}$ & $3.10_{-1.30}^{+1.03}$ & $<7.38$ & Sage07 \\
NGC~741 & $2.70_{-1.83}^{+2.19}$ & - & $<8.28$ & Evans05 \\ 
NGC~1399 & $2.31_{-0.31}^{+0.31}$ & $1.30_{-0.04}^{+0.03}$ & 7.47 & Prandoni10 \\
NGC~1407 & $2.70_{-1.53}^{+1.83}$ & - & $<7.53$ & Sage07 \\
NGC~4365 & $2.17_{-8.32}^{+1.70}$ & - & $<7.62$ & Young11 \\
NGC~4406 & $1.60_{-0.41}^{+0.37}$ & $1.33_{-0.05}^{+0.05}$ & $<7.40$ & Young11 \\
NGC~4472 & $2.06_{-0.27}^{+0.25}$ & $2.06_{-0.06}^{+0.06}$ & $<7.25$ & Young11 \\
NGC~4552 & $1.45_{-0.48}^{+0.45}$ & $3.00_{-0.45}^{+0.45}$ & $<7.28$ & Young11 \\
NGC~4636 & $1.51_{-0.16}^{+0.11}$ & $1.59_{-0.04}^{+0.04}$ & $<6.87$ & Young11 \\
NGC~4649 & $2.85_{-0.60}^{+0.56}$ & $2.40_{-0.13}^{+0.12}$ & $<7.44$ & Young11 \\
NGC~5846 & $5.00_{-3.25}^{+3.25}$ & - & $<7.78$ & Young11 \\
\hline \hline
\end{tabular}\label{table:EarlyTypeCOAbun}
\end{center}
\vspace{-0.0in} \footnotesize Fe/O abundance ratio of hot gas and molecular gas mass ($M_{\rm H_2}$) of NGC~5866 and the elliptical galaxies from \citet{Humphrey06} and \citet{Ji09}. For galaxies included in both \citet{Humphrey06} and \citet{Ji09}, we adopt \citet{Ji09}'s measurements as they use better X-ray data. \citet{Ji09} perform X-ray measurements based on different instruments (\emph{Chandra} ACIS, \emph{XMM-Newton} EPIC and RGS) and with different aperture sizes (diameters of $1^\prime$ and $4^\prime$). We herein adopt the \emph{XMM-Newton} EPIC $1^\prime$-diameter measurements, which should be the most similar as our measurement of NGC~5866, and is less affected by the clustered environment of some galaxies. The Fe/O ratio used in the plots are summarized in the ``Fe/O'' column. We also list \citet{Ji09}'s \emph{XMM-Newton} EPIC $4^\prime$-diameter measurements in the ``Fe/O ($4^\prime$)'' column for comparison. The differences in these two columns do not affect our results discussed in this paper. Errors of the Fe/O ratio are quoted at 90\% confidence level in this table, but are plotted at $1~\sigma$ confidence level in Figs.~\ref{fig:FeORatioTC} and \ref{fig:FeORatioMH2} to clarify. The molecular gas mass $M_{\rm H_2}$ is based on CO observations, with references listed in the last column. Evans05: \citet{Evans05}; Sage07: \citet{Sage07}; Prandoni10: \citet{Prandoni10}; Young11: \citet{Young11}.
\end{table}

The temperature of the hot gas in the inner halo seems too low to be primarily heated by SNe. In gas-poor galactic bulges like NGC~5866, hot gas is the dominant phase and the radiative cooling of the shocked ISM is overall not important in the SN energy budget (e.g., \citealt{LiM15}). The specific enthalpy of old stellar feedback (mass loss of evolved low-mass stars and Type~Ia SNe heating) is $\sim4\rm~keV$ per particle. Such a specific enthalpy gives an average gas temperature of $\sim1\rm~keV$, considering the energy equipartition between the hot gas thermal energy and the energy stored in magnetic field and cosmic rays as well as turbulent motion \citep{Li09,Tang09,Tang10}. On the other hand, adopting a central velocity dispersion of $\sigma_v\sim162\rm~km~s^{-1}$ obtained from the HyperLeda database, the expected temperature of gas thermalized in the gravitational potential of NGC~5866 would be $\sim0.2\rm~keV$, quite close to the measured hot gas temperature of $\sim0.25\rm~keV$ in the inner halo (Table~\ref{table:SpecFitResult}). Therefore, the low measured hot gas temperature \emph{apparently} indicates that any additional sources of non-gravitational heating, such as Type~Ia SNe, play a relatively minor role (contributes only $\sim20\%$ of the total thermal energy). 

Although located in a very low density environment (the environmental galaxy number density of NGC~5866 is $\rho=0.24$; in comparison, $\rho\geq0.6$ is classified as clustered in \citealt{Li13a}), it is still possible that there exists a low-density intra-group medium around NGC~5866. However, neither of the existing \emph{Suzaku} or \emph{Chandra} data as presented in this paper is deep enough to probe the hot gas distribution at large radii. The \emph{Chandra} X-ray intensity profiles at small radii, approximately within the spectral analysis regions of the present paper, are presented in \citet{Li09,Li13a}. These profiles do not reveal any diffuse soft X-ray emission above the stellar contributions at $\gtrsim1^\prime$ (or roughly the inner halo region in this paper; Fig.~\ref{fig:image}b). Therefore, based on the existing X-ray data, we do not find significant evidence for the existence of an intra-group medium, which, if presents, could help to raise the hot gas temperature in the outer halo (e.g., \citealt{Sun09}). 

It is suggested that the X-ray emissivity of early-type galactic coronae, and also the fraction of SN ejecta detected in soft X-ray, are highly affected by the specific energy of the hot gas (energy per unit mass, or temperature if the gas is thermalized; \citealt{Li11}). If this specific energy is reduced by some processes, there will be a lower temperature of the hot gas and slower galactic outflow. As a result, the hot gas density and X-ray emissivity will be raised and there will be a larger amount of heavy elements produced by Type~Ia SNe (such as Fe) detected in soft X-ray. Cool-hot gas interaction, as indicated by the relatively large amount of cool gas (Table~\ref{table:EarlyTypeCOAbun}; Fig.~\ref{fig:FeORatioMH2}) and the presence of extraplanar dusty filaments in NGC~5866 (Fig.~\ref{fig:HSTfilaments}), could provide a natural explanation of both the low temperature and high Fe abundance of the hot gas. 

The apparently low Type~Ia SNe contribution to the heating of the hot gas is very unlikely, because NGC~5866 shows so many features related to Type~Ia SNe feedback, such as the Fe enrichment detected in hot gas and the consistency of the gravitational energy of many extraplanar dusty filaments with a single SN energy \citep{Li09}. Furthermore, the significant raising of the hot gas temperature in the outer halo (Fig.~\ref{fig:kTcompare}; \S\ref{subsec:SpecAnalysis}) in such a low-mass isolated galaxy is quite uncommon (e.g., \citealt{Fukazawa06}). Although we cannot absolutely exclude the contribution from a low-density intra-group medium based on the existing \emph{Suzaku} or \emph{Chandra} data, it is natural that there exists some additional cooling processes, such as the cool-hot gas interaction. The presence of cool gas could affect the surrounding hot gas properties via mass loading and enhanced radiative cooling, both processes reduce the specific energy and so the temperature of the hot gas. 

The unusually strong cool-hot gas interaction in NGC~5866 as an S0 galaxy could also help to explain the high Fe abundance detected in soft X-ray. The cool gas is often less metal-enriched, and the mass loading of cool gas into hot phase will result in a lower average metallicity of the hot gas. However, most of the SN ejecta contained in the high-temperature (typically $\gtrsim1\rm~keV$) hot gas has too low emissivity to be detected in soft X-ray, because the hot gas density of gas-poor galactic bulge is too low (typically $\lesssim10^{-4}\rm~cm^{-3}$, e.g., \citealt{Tang10}). Therefore, the cool-hot gas interaction will have an opposite effect on the observed gas metallicity, by significantly increasing the soft X-ray emissivity and so the \emph{emission-measure-weighted} metallcity as obtained from X-ray observations. Such a difference between the emission-measure- and mass-weighted metallicity is also evidenced by numerical simulations (e.g., \citealt{Tang09,Crain13}), indicating that we are missing from the detection in X-ray observations of a large fraction of gas-phase metals, as also being estimated in \S\ref{subsec:HotGasMetallicity}. 

\section{Summary and Conclusion}\label{sec:Summary}

We present detail calibration of the \emph{Suzaku} observation of an isolated S0 galaxy NGC~5866. By jointly analyzing this \emph{Suzaku} data together with an archival \emph{Chandra} observation, we obtain the hot gas temperature in the inner ($r\leq1^\prime$) and outer ($1^\prime<r\leq1.6^\prime$) halos, as well as the O and Fe abundances in the inner halo. These measurements of the hot gas properties are much more accurate than previous measurements with the \emph{Chandra} data only, allowing for a comparison with other galaxies.

We find a significant supersolar Fe/O abundance ratio of $\rm Fe/O=7.63_{-5.52}^{+7.28}$ at $90\%$ statistical confidence level in the inner halo of NGC~5866, indicating a clear metal enrichment from Type~Ia SNe. The Fe/O abundance ratio of NGC~5866 is also significantly higher than other disk (spiral+S0) galaxies, and even many of the elliptical galaxies. We also find that NGC~5866 simultaneously has the highest Fe/O abundance ratio and molecular gas mass ($M_{\rm H_2}$) among a small sample of gas-poor early-type galaxies, although the poor and mostly upper limit measurement of $M_{\rm H_2}$ prevents us from claiming a clear positive correlation between Fe/O and $M_{\rm H_2}$. The total amount of Fe detected in the inner halo of NGC~5866 is $\sim2\times10^5f^{1/2}\rm~M_\odot$. In contrast, the total amount of Fe injected by Type~Ia SNe within the dynamical cross timescale of the galactic outflow is $\sim2\times10^4\rm~M_\odot$, while that within the age of the galaxy since the last starburst is $\sim10^7\rm~M_\odot$. Considering the uncertainty of the volume filling factor $f$ which cannot be too low in this non-starburst S0 galaxy, we could conclude that NGC~5866 could preserve a larger than usual fraction, but far from the total amount of Fe injected by Type~Ia SNe.

NGC~5866 exhibits a lot of extraplanar dusty filaments extending perpendicular to the galactic disk within the inner halo. These filaments indicate the existence of interaction between SNe and cool gas, which probably triggers cool gas mass loading into the hot phase and enhances the radiative cooling. We find clearly lower hot gas temperature in the inner halo than in the outer halo, which is apparently a result of this cool-hot gas interaction, if this well isolated galaxy is not highly affected by a high-temperature intra-group medium. This low temperature in the inner halo is also roughly consistent with gravitational heating of the material lost by evolved stars, \emph{apparently} leaving little room for additional non-gravitational heating such as by Type~Ia SNe. However, heating by Type~Ia SNe is very likely to be important in this galaxy showing clear Fe enrichment and cool-hot gas interaction. Therefore, additional cooling processes must exist. 

The low temperature and high Fe abundance of the hot gas in the inner halo of NGC~5866 strongly suggest that the presence of cool gas in gas-poor galaxy bulges could trigger cool-hot gas interaction, resulting in a lower hot gas temperature and a larger fraction of SN ejecta detected in soft X-ray around the galaxy. However, most of the metal-enriched SNe ejecta is still missing from a clear detection in X-ray.

\acknowledgements

Li Jiang-Tao would like to acknowledge Roberts Shawn R., Wang Q. Daniel, and the \emph{Suzaku} helpdesk for discussions on the \emph{Suzaku} background and also the anonymous referee for very helpful comments. This work is supported by NASA through the grant NNX13AE87G and NNX15AM93G.

\end{document}